\documentclass[acmsmall, anonymous=False, review=False]{acmart}
\usepackage{amsmath,array,multirow,graphicx,pdflscape}
\usepackage{bm} 
\usepackage[utf8]{inputenc}

\AtBeginDocument{%
  \providecommand\BibTeX{{%
    \normalfont B\kern-0.5em{\scshape i\kern-0.25em b}\kern-0.8em\TeX}}}

\setcopyright{rightsretained}
\acmJournal{PACMHCI}
\acmYear{2022} \acmVolume{6} \acmNumber{CSCW2} \acmArticle{333} \acmMonth{11} \acmPrice{}\acmDOI{10.1145/3555225}

\begin{document}

\title[The Risks, Benefits, and Consequences of Prepublication Moderation]{The Risks, Benefits, and Consequences of Prepublication Moderation: Evidence from 17 Wikipedia Language Editions}

\author{Chau Tran}
\affiliation{%
  \institution{New York University}
  \city{New York, NY}
  \country{United States}
}
\email{chau.tran@nyu.edu}

\author{Kaylea Champion}
\affiliation{%
  \institution{University of Washington}
  \city{Seattle, WA}
  \country{United States}
}
\email{kaylea@uw.edu}

\author{Benjamin Mako Hill}
\affiliation{%
  \institution{University of Washington}
  \city{Seattle, WA}
  \country{United States}
}
\email{makohill@uw.edu}

\author{Rachel Greenstadt}
\affiliation{%
  \institution{New York University}
  \city{New York, NY}
  \country{United States}
}
\email{greenstadt@nyu.edu}

\renewcommand{\shortauthors}{Tran et al.}


\begin{abstract}
Many online communities rely on postpublication moderation where contributors---even those that are perceived as being risky---are allowed to publish material immediately and where moderation takes place after the fact. An alternative arrangement involves moderating content before publication. A range of communities have argued against prepublication moderation by suggesting that it makes contributing less enjoyable for new members and that it will distract established community members with extra moderation work. We present an empirical analysis of the effects of a prepublication moderation system called \textit{FlaggedRevs} that was deployed by several Wikipedia language editions. We used panel data from 17 large Wikipedia editions to test a series of hypotheses related to the effect of the system on activity levels and contribution quality. We found that the system was very effective at keeping low-quality contributions from ever becoming visible. Although there is some evidence that the system discouraged participation among users without accounts, our analysis suggests that the system's effects on contribution volume and quality were moderate at most. Our findings imply that concerns regarding the major negative effects of prepublication moderation systems on contribution quality and project productivity may be overstated.
\end{abstract}

\maketitle
\section{Introduction}

Successful commons-based peer production communities struggle to maintain the quality of the knowledge bases that they construct in spite of vandalism, trolls, spam, and harassment~\cite{kraut2012building, 10.1145/3359157, 10.1145/3134659}. These threats, if left unchecked, can have a ripple effect: toxic behaviors elicit more toxic behaviors \cite{10.1145/3415179, anderson2018toxic}, which leads to increased rates of newcomer rejection, and ultimately, to a decreased contributor base~\cite{doi:10.1177/0002764212469365, 10.1145/2641580.2641614, 10.1145/1099203.1099206}.
Site administrators therefore often employ content moderation to censor harmful content~\cite{gillespie2018custodians}.
What kind of moderation system should communities employ? Should content be moderated before it is made public, denying good faith contributors a sense of immediacy, or should content only be moderated after publication? 

For knowledge base websites like Wikipedia, Quora, and Fandom (formerly known as Wikia), administrators must find a balance between allowing content to be speedily updated and keeping inappropriate content from being displayed to the public. Although screening a large volume of content in real time might not be feasible for all communities, some sites use a prepublication moderation system in combination with postpublication review. Prepublication moderation can be applied to a subset of contributions, such as those from untrusted users. Although there is some agreement that additional quality assurance mechanisms can be helpful in fighting vandalism, there is deep concern regarding the unforeseen impact of these systems. 
In the case of Wikipedia, quality control measures have been described as the cause of the platform's decline in growth~\cite{doi:10.1177/0002764212469365}. Does prepublication moderation negatively impact the volume of content submitted by less trusted users or decrease the likelihood that they will continue to contribute? Does a change in workflow that increases the level of prepublication moderation tend to inspire more or better contributions? 

In this paper, we present a quantitative evaluation of a prepublication moderation system called \textit{FlaggedRevs} and its deployment in 17 Wikipedia communities. By collecting and processing data available from these wikis, we used a community-level panel data interrupted time series (ITS) analysis, as well as a user-level general linear mixed model (GLMM), to identify the effects of FlaggedRevs on several different outcomes. 
Our work makes a series of empirical contributions to the study of peer production systems by testing a set of hypotheses drawn from prior work and conversations within online communities and through an evaluation of a prepublication review system that is similar to those used in a large range of other settings. First, we find that the system is very effective at blocking low-quality contributions from ever being visible. In analyzing its side effects, we found, contrary to expectations and most of our hypotheses, little evidence that the system neither raises transaction costs sufficiently to inhibit participation by the community as a whole, nor measurably improves the quality of contributions.

\section{Background}

\subsection{Peer Production and Barriers to Entry}
Peer production describes a widespread and influential type of online collaborative work that involves the mass aggregation of many small contributions from diversely motivated individuals working together over the Internet \cite{benkler_peer_2015}. Although the most famous examples of peer production include free/libre open source software and Wikipedia, peer production also describes collaborative filtering on sites like Reddit, Q\&A on sites like Quora, and activity in a range of other digital knowledge bases and communities. 

Theorized first by Benkler in 2002 \cite{benkler_coases_2002, benkler2006wealth}, peer production is described as being made possible through extremely low transaction costs. In other words, it is possible because it is extremely easy to join peer production communities and make contributions, relative to markets and firms. 
Broader theories of public goods and collective action support the idea that low participation costs help communities achieve critical mass \cite{marwell1993critical}, facilitate fluid or open community boundaries \cite{olson2012logic}, and attract a greater diversity of perspectives \cite{benkler2006wealth, bennett_logic_2012, kane_emergent}.  

Empirical work suggests that low barriers to entry also helps stimulate participation across users with and without accounts \cite{antin2010reader}, and can introduce diverse perspectives from a wider pool of participants~\cite{anthony2009reputation, kane_emergent}. 
By contrast, high barriers to entry can deter participation \cite{oberschall1973social, olson1989collective}.
Although Friedman and Resnick \cite{friedman2001social} argued that low barriers come at a substantial social cost to communities and it may be advantageous to expect newcomers to ``pay their dues,'' Hill and Shaw 
\cite{doi:10.1177/0093650220910345} found that raising barriers to entry in subtle ways can ultimately decrease contributions both from newcomers and more established participants.

A range of social computing studies have found evidence that quality control policies tend to negatively affect the growth of a peer production community because new users are particularly sensitive to the feedback that they receive for their good-faith, but often low-quality, efforts ~\cite{geiger2012defense, halfaker_newbie, doi:10.1177/0002764212469365, doi:10.1177/0093650220910345}.
Attracting and retaining new users are necessary for peer production communities to maintain the resources that they have built and to continue to grow~\cite{halfaker_newbie}. Contributors seek their peers' recognition for their work~\cite{10.2307/20141769, 10.1111/j.1083-6101.2008.00416.x}, and 
prior work has observed that reducing the immediate visibility of contributions might diminish contributors' enthusiasm and long-term commitment \cite{greis2014can}.

\subsection{Moderation and Community Health}
Prior research on peer production communities has shown that malicious actions such as trolling, spamming, and flaming can undermine a community’s purpose and can drive members away \cite{kraut2012building, lea1992flaming, cruz2018trolling, leavitt2015throwaway}. 
Conversely, a ``healthy'' environment can promote continued participation from existing contributors and can motivate new participants to join \cite{nonnecke2001lurkers}. For example, a study by Wise et al.~shows that ``a moderated online community elicited greater intent to participate than an unmoderated community'' \cite{moderation_response_rate}.
To limit the negative impact of antisocial content and to create healthy communities, nearly all online communities engage in some type of content moderation using either professional or volunteer moderators \cite{gillespie2018custodians}. In many cases, platforms are obligated to moderate sites to some extent, because of local norms and laws \cite{jiang_understanding_2021}. 

CSCW scholars have recently turned their attention to content moderation. Although we will not attempt to provide full accounting of that scholarship, \citeauthor{kiene_volunteer_2019}'s 2019 CSCW panel abstract \citep{kiene_volunteer_2019} and \citeauthor{seering_reconsidering_2020}'s 2020 review article \citep{seering_reconsidering_2020}  are two useful entry points.
One key finding from this research is that moderating problematic content is a burden on moderators. Moderators, who are frequently volunteers, engage in emotional labor as they are forced to confront toxic material and negotiate their own emotional responses, perhaps while simultaneously negotiating between parties and seeking to nurture a positive environment where all members feel engaged, empowered, and cared for~\cite{10.1145/3290605.3300372, gillespie2018custodians, gillespie2020content, doi:10.1177/1461444820964968}. Moderation techniques that ultimately discourage future toxic behavior may diminish this emotional labor, prevent burnout and support moderators~\cite{dosono2019moderation}.

\subsection{Moderation Design Decisions}

In an influential article on content moderation, \citet{grimmelmann2015virtues} describes the design of a moderation system in terms of four key characteristics: (1) \textit{human} vs. \textit{computer}, (2) \textit{secret} vs. \textit{transparent}, (3) \textit{ex ante} vs. \textit{ex post}, and (4) \textit{centralized} vs. \textit{distributed} \cite{grimmelmann2015virtues}. The first characteristic captures whether moderation relies on human moderators or whether it is partially or entirely assisted by automated software. The second reflects the choice of whether to be transparent with rules and guidelines in order to educate users with clear feedback.  Research has shown that transparency can have an impact in keeping members engaged while simultaneously enforcing rules and norms~\cite{10.1145/3359252, 10.1145/3359294} whereas secrecy minimizes the risk that rules will be circumvented by bad actors. The third feature describes how moderators can choose whether to act \textit{ex ante} (i.e., taking a preventive action against contributions deemed inappropriate and antinormative) or \textit{ex post} (i.e., updating or removing harmful content after it has been posted). Fourth and finally, a moderation system can be either centralized by having a dedicated group of moderators who watch over the entire community, or decentralized/distributed by having members within the community watching over each other.
In Grimmelmann's treatment, all four dimensions are orthogonal. For example, a community might employ automated secret centralized moderation either \textit{ex post} or \textit{ex ante}.

In this work, we focus on the key distinction between \textit{prepublication moderation} and \textit{postpublication moderation} that corresponds directly to Grimmelmann's third dimension of \textit{ex ante} and \textit{ex post} moderation. We diverge from Grimmelmann by using terminology drawn from Coutino and José for the same concept~\cite{7975786}. In postpublication moderation (i.e., \textit{ex post} or ``post-moderation'') content contributed by users is made instantly visible online and remains there until it is changed or removed \cite{grimmelmann2015virtues, 7975786}. Because inappropriate content is examined and removed or updated after the fact, postpublication moderation allows real-time interactions between users. 
Postpublication moderation has also been described as particularly effective at eliciting contributions because contributors see their contributions appear immediately which can lead to a sense of belonging and self-efficacy as well as faster collaboration~\cite{grimes-viort_2010, moderation_response_rate}. Indeed, real-time communication and interaction have been shown to increase social connections between peers and further engagement in the community \cite{nowak2005influence}. Finally, when contributions are immediately reflected and new information can be updated instantly, it may further incentivize writers to continue contributing to keep information updated~\cite{wahl2010audience}. 

Postpublication moderation techniques are often favored among large social media sites, discussion boards, forums, and similar, for facilitating seamless communication between users and reducing the risk of congestion \cite{grimmelmann2015virtues}.
However, postpublication moderation has several disadvantages. First, it can be ineffective in fighting vandalism in that the damage done is always visible---even if only briefly. Second, the reactive nature of postpublication moderation systems means that malicious behaviors such as trolling, flaming, and spamming can potentially go unnoticed, leading to low satisfaction among users~\cite{butler1999group, diakopoulos2011towards}. 

A prepublication moderation approach (i.e., \textit{ex ante} or ``pre-moderation'') is one that funnels contributions through an inspection and approval process conducted by human moderators, algorithms, or some combination of the two \cite{grimmelmann2015virtues, 7975786}. 
This means that contributors must wait for their work to be explicitly approved by a moderator before becoming visible to the general public. The most obvious benefit of prepublication moderation is that harmful content is more likely to be rejected before it affects other users. Additionally, prepublication moderation can help members internalize normative behaviors through trial and error. Grimmelmann argues that the feedback members of the community receive for their contributions will help them ``learn to check themselves'' before they submit their next contribution \cite{grimmelmann2015virtues}. 

Although prepublication systems can mitigate harmful content, they have several drawbacks. First, the effort and time required to vet content may be a substantial burden on limited moderator resources \citep{gillespie2020content}. Time spent on the work of reviewing might mean that moderators have less time to spend contributing to the community in other ways. 
Furthermore, waiting for an extended amount of time for contributions to be explicitly approved creates ``a lack of instant gratification on the part of the participant'' which may reduce individuals' desire to contribute again and result in less productive newcomers \cite{veglis2014moderation, 10.1145/2641580.2641614}. 

For communities whose mission is to provide an environment with minimal risk of harassment and abuse, and deliver high-quality and accurate content (i.e., journalistic blogs, some types of mailing lists, and some Q\&A sites), or where the legal consequences of publishing problematic material are very high, prepublication moderation has been described as an ideal solution \cite{10.1007/978-3-319-07632-4_13, veglis2014moderation}.  
For example, ``moderated'' mailing lists where messages must be reviewed by an administrator before being sent, have been widespread for decades.\footnote{e.g., \url{https://www.lsoft.com/manuals/16.0/htmlhelp/list\%20owners/MailingListsTypes.html}} Similarly, edits to existing content on StackExchange by new users are vetted before being posted. Many widely read blogs and news websites allow for the submission of links, comments, and opinion articles that are vetted before being published. On a personal communications level, Squadbox allows users to create a ``squad'' of friend moderators to filter out harassing messages before they are seen by a recipient~\cite{mahar2018squadbox}.
Of course, pre- and postpublication moderation are not mutually exclusive, and most systems that do the former will also do at least some of the latter. The work of moderation often involves both attempts to prevent and react to problematic material~\citep{grimmelmann2015virtues, seering_reconsidering_2020, 7975786}.


\subsection{Testing the Effects of Prepublication Moderation on Contribution Actions}

Prepublication moderation systems are designed to proactively protect a platform or community by preventing damaging content from being displayed publicly \cite{grimmelmann2015virtues, klonick2017new}. Although our study is primarily interested in the unintended side effects of prepublication moderation, evaluating a system fully means that we must first understand whether it achieves its primary goal. We assess whether a prepublication moderation system is indeed functioning as intended by offering a single hypothesis in three parts: \textit{prepublication moderation will be associated with a smaller number of low-quality contribution actions from affected users that are made visible to the public (H1a)}. Because the work of unaffected users are not subject to review, we expect that \textit{prepublication moderation will not be associated with a difference in the number of low-quality contribution actions from unaffected users that are made visible to the public (H1b)}---i.e., we expect a null effect. Finally, we also expect to see an overall reduction in the number of low-quality contribution actions that are made public. In other words, \textit{prepublication moderation will be associated with a smaller number of low-quality contribution actions from the community overall that are made visible to the public (H1c)}.

In a related sense, we aim to examine the argument by Grimmelmann that states that prepublication moderation enforces the norms of the community by making its members ``think twice'' before posting \cite{grimmelmann2015virtues}. In this case, the explicit approval or disapproval that a user receives will serve as constructive feedback for future submissions, while simultaneously making users more conscious of the fact that their work is being actively assessed. This reflects the insight that proactive measures of content moderation and production control can play an important role in encouraging prosocial behavior \cite{10.1145/2998181.2998277, 10.1145/3311957.3359478, zhu2012organizing, 10.1145/3311957.3361855}. We thereby anticipate that \textit{prepublication moderation will be associated with higher contribution quality from affected users (H2a)}. Once again, we also expect a null effect for the unaffected user group, stating that \textit{prepublication moderation will not significantly affect the contribution quality of unaffected users (H2b)}. Finally, we also expect that \textit{prepublication moderation will be associated with higher contribution quality of the community overall (H2c)}.

Finally, we turn to research that has shown that additional quality control policies may negatively affect the growth of a peer production community \cite{halfaker_newbie, doi:10.1177/0002764212469365, 10.1145/2641580.2641614}. 
Our third set of hypotheses anticipates that the benefits described in H1 and H2 come with a trade-off related to the types of unanticipated negative side effects shown in the evaluation of other moderation systems. Because of the reduced sense of self-efficacy associated with contributing, we anticipate that \textit{prepublication moderation systems will  be associated with fewer contribution actions from affected users (H3a)}. Although the intuition is less obvious, we are mindful of previous research that suggests that a prepublication system in a community-moderated environment will also negatively affect the contribution rate of established users whose contribution are not directly subject to review \cite{halfaker_newbie}. Such systems require effort from experienced contributors and may result in a net increase in the demands on these users' time. Therefore, we hypothesize that \textit{prepublication moderation systems will be associated with fewer contribution actions from unaffected users (H3b)}. Because both our previous hypotheses point to reduced contribution rates, we also suggest that \textit{prepublication moderation systems will be associated with reduced contribution actions from members of the community overall (H3c)}.

The higher barrier to entry posed by prepublication review, combined with the delayed intrinsic reward, might be disheartening enough to drive newcomers away \cite{10.1145/2641580.2641614}. Different from our previous hypotheses, our fourth hypothesis focuses solely on new users. We hypothesize that \textit{the deployment of a prepublication moderation system will negatively affect the return rate of newcomers (H4)}.

\section{Empirical Setting}

As one of the largest peer production websites on the Internet, Wikipedia describes itself as ``the free encyclopedia that anyone can edit'' because the site allows editing by the general public with very few barriers.\footnote{\url{https://www.wikipedia.org}} Visitors can edit the vast majority of Wikipedia pages without creating an account, providing an email address, or establishing a persistent identity. Wikipedia exists in more than 300 languages, and each language edition has its own community with its own rules and administrative structure.
Unfortunately, the ease of editing attracts malicious and bad-faith participants~\cite{beyond_vandalism_Pnina}. Wikipedia relies on transparency to defend against problematic contributions. Every contributed revision is recorded, including the state of the article before and after the revision, and is made available for public review. Low-quality revisions found to violate community policies or norms may be removed and any article can easily be reverted to any previous version. 
Although larger Wikipedia language editions like English may be able to field sufficient volunteer resources to fight vandalism, and even to build, customize, and test advanced automated antivandalism tools \cite{geiger2013levee, geiger_work_2010}, volunteers in smaller communities may not have the same resources. Although automated tools to help protect Wikipedia are an active area of social computing research, the task is far from complete~\cite{halfaker_ores_2020}. Uncaught vandalism, hoaxes, and disinformation frequently remain hidden for extended periods of time and detrimentally affect the community's reputation and sustainability~\cite{kumar_disinformation_2016}.

To address risks associated with unmoderated content becoming visible, the Wikimedia Foundation and the German Wikimedia chapter (Wikimedia Deutschland) collaborated to develop \textit{Flagged Revisions} (\textit{FlaggedRevs} for short), a highly configurable extension to the MediaWiki software that runs Wikipedia.\footnote{\url{https://meta.wikimedia.org/wiki/Flagged_Revisions}} FlaggedRevs is a prepublication content moderation system in that it will display the most recent ``flagged'' revision of any page for which FlaggedRevs is enabled instead of the most recent revision in general.
FlaggedRevs is designed to ``give additional information regarding quality,'' by ensuring that revisions from less-trusted users are vetted for vandalism or substandard content (e.g., obvious mistakes because of sloppy editing) before being flagged and made public. The FlaggedRevs system also displays the moderation status of the contribution to readers.\footnote{\url{https://meta.wikimedia.org/wiki/Flagged_Revisions}}
We look to Wikipedia for the same reasons many other social computing scholars have. First, Wikipedia is globally significant as an information source and a successful site of peer production. Further, the Wikimedia Foundation makes incredibly detailed research data publicly available. In our particular case, Wikipedia deployed FlaggedRevs with a remarkably high degree of transparency that made our study possible. 

Although there are many details that can vary based on the way that the system is configured, FlaggedRevs has typically been deployed in the following way on Wikipedia language editions. First, users are divided into groups of trusted and untrusted users. Untrusted users typically include all users without accounts as well as users who have created accounts recently and/or contributed very little. Although editors without accounts remain untrusted indefinitely, editors with accounts are automatically promoted to trusted status when they clear certain thresholds determined by each language community. For example, German Wikipedia automatically promotes editors with accounts who have contributed at least 300 revisions accompanied by at least 30 comments.\footnote{https://noc.wikimedia.org/conf/highlight.php?file=flaggedrevs.php} 

Contributions to articles made by trusted users are automatically made public (i.e., ``flagged'')---just as they would be if FlaggedRevs were not deployed. Contributions to articles made by untrusted users are not made visible immediately but are instead marked as provisional and placed into a queue where they wait to be reviewed and approved. The moderation process is conducted by trusted volunteers with review privileges who can flag revisions outright, reject the proposed revision by reverting it, or edit the proposed revision and publish it.
The FlaggedRevs extension must be installed on a per-wiki basis, which means that it is deployed in some Wikipedia language editions, but not in most. New deployments of FlaggedRevs were placed under a moratorium in April 2017 because of technical problems and the high staffing costs associated with configuring and maintaining the system.\footnote{ibid.} Prior to this moratorium, 24 language editions of Wikipedia enabled FlaggedRevs.\footnote{Although it did not use the system, English Wikipedia implemented a similar vetting system called \textit{Pending Changes}, but on a much smaller scale.} Additionally, some other Wikimedia projects (e.g., Wiktionary, Wikibooks, Wikinews, Wikiquote, and Wikisource) have also deployed the system. 

Despite its importance and deployment in a number of large Wikipedia communities, very little is known regarding the effectiveness of the system and its impact. A report made by the members of the Wikimedia Foundation in 2008 gave a brief overview of the extension, its capabilities and deployment status at the time, but acknowledged that ``it is not yet fully understood what the impact of the implementation of FlaggedRevs has been on the number of contributions by new users.''\footnote{\url{https://meta.wikimedia.org/wiki/FlaggedRevs_Report_December_2008}} Our work seeks to address this empirical gap.

\section{Methods}

\subsection{Data}

We first collected an exhaustive dataset of revision and content moderation activities that occurred on all Wikipedia language editions that enabled FlaggedRevs. This included 24 wikis in total. The datasets that we collected are publicly available in the Wikimedia Downloads website.\footnote{\url{https://dumps.wikimedia.org/}} For each wiki, we downloaded the \texttt{stub-meta-history} XML dumps that contained the metadata for all revisions made to pages that were public (i.e., not deleted) at the time that the database dump was created. We also used the \texttt{flaggedrevs} SQL dump that contained information regarding all revisions that were reviewed under the FlaggedRevs system in each wiki. 
There is substantial variation across these wikis, including language, activity level, and organizational structure. Wikis vary enormously in size, with numbers of contributions ranging from thousands to millions each year. Furthermore, the way each wiki configured FlaggedRevs also varies. Differences include the criteria for a revision to be flagged, guidelines for determining whether the quality of a revision warrants publication, and differences in the threshold for treating contributors as trusted.
We used a series of custom software scripts to convert the data in both database dumps into a tabular format, which allowed us to build measures associated with the concepts in our hypotheses. We have placed the full code and datasets used for these analyses into an archival repository in the Harvard Dataverse.

Because of the way that we constructed our hypotheses, we built two datasets with different metadata and units of analysis. The first dataset is a \textit{wiki-level} dataset that is used to test H1, H2, and H3. The second dataset is a \textit{user-level} dataset that is used to test H4. Each row of this second dataset represents an individual user in one of the wikis in our analysis.

In the wiki-level dataset, our unit of analysis is the wiki month. We constructed these data by starting with the raw dataset collected initially where each row involves a revision to a page in each language's Wikipedia. We then proceeded to restrict our dataset to only article pages by excluding revisions to pages in nonarticle ``namespaces'' (i.e., discussion pages, special pages).\footnote{\url{https://en.wikipedia.org/wiki/Wikipedia:Namespace}} We did so because Wikipedia communities have a different set of guidelines when reviewing non-article contributions for vandalism and because FlaggedRevs is typically not enabled for these pages. 
Next, we grouped revisions by type of user in ways that correspond to our subhypotheses as described in §\ref{sec:outcomes}. We then aggregated the data and counted the total number of contributions made by each editor group by month for each wiki. We excluded any wikis with fewer than 30 new contributions per month, on average, that are made by each of our editor groups.

For each published contribution, we must know the time that the contribution was made, the time that it was published (i.e., flagged), and the manner it was published (either manually or automatically).  Because they are critical for our analyses, we omitted any wiki for which we could not obtain these data. As a result, our empirical setting is a population of 17 Wikipedia language communities: Albanian, Arabic, Belarusian, Bengali, Bosnian, Esperanto, Persian, Finnish, Georgian, German, Hungarian, Indonesian, Interlingua, Macedonian, Polish, Russian, and Turkish. 

Because each wiki enabled FlaggedRevs at a different point in time, and because we only sought to estimate the immediate impacts of the intervention, our datasets are restricted to revisions made in the 12 month period before and after FlaggedRevs was enabled. Finally, because these wikis vary vastly in size, we transformed our outcome measures for H1, H2, and H3 into standard deviation units within each wiki. This ensures that the outcome measure from each wiki has an equally weighted impact on our analysis, regardless of the wiki's size or average activity level.
Standardizing activity levels across wikis in this way is also helpful in ensuring that the parametric assumptions of our regression models are met.


The user-level dataset does not have additional exclusion criteria but only contains information regarding new editors from the 17 wikis listed above. We consider each unique editor ID (either an account name or an IP address) to be a unique user. Because each row in the dataset contains information on an editor, our sample---and statistical power by extension---is much larger. As a result, we restrict our dataset to users whose first contributions are made within the period 90 days before and after FlaggedRevs is enabled. There are 1,972,861 observations in the user-level dataset. 

\subsection{Outcome Measures}
\label{sec:outcomes}

Our H1 hypotheses pertain to the number of low-quality contributions that are made visible to the public. We operationalize this as the \textit{number of visible rejected contribution actions} that reflects the aggregated and standardized number of visible reverted contributions per month for each wiki. When a contribution action to an article is rejected, the editor acting as a moderator typically performs a \textit{revert} action to nullify any changes that a contribution has made. With FlaggedRevs, edits are never ``disapproved'' explicitly. Instead, pages are simply reverted to a previously accepted version. Before FlaggedRevs is enabled on a wiki, all contributions are instantly accessible by the public, even if they are later reverted. After FlaggedRevs is enabled, contributions made by the affected editor groups have to wait to be approved (flagged) before becoming visible. Readers of the affected article see the last approved version throughout this process. 
    
Our H2 hypotheses suggest that prepublication review will affect the quality of contributions overall. We operationalize quality in two ways. First, we use the \textit{number of rejected contributions} that we operationalize as the number of reverts. We follow other Wikipedia researchers by measuring reverts as any contribution that returns a page to a previous state and subjected edits that are undone by a revert as reverted.  Because reverting is the most frequent approach used to fight vandalism \cite{Kittur:2007:HSS:1240624.1240698}, the number of reverted contributions is a fairly reliable indicator of whether the contributions received are of poor quality or otherwise unwelcome. A large body of previous research into Wikipedia has used reverts as a measure of quality \cite{Kittur:2007:HSS:1240624.1240698, chau_tor_ieee,bongwon_slow_growth}.  We also test our second hypotheses using average quality that we operationalize as \textit{revert rate}. Following previous research~\cite{chau_tor_ieee, bongwon_slow_growth}, revert rate is measured as the proportion of contributions that are eventually reverted. Together with the number of reverted contributions, this measure offers a more complete picture of quality.

Our H3 hypotheses call for a measure of user productivity. Although there are a range of ways to measure productivity, we follow Hill and Shaw \cite{doi:10.1177/0093650220910345} and a range of other scholars who operationalize wiki-level productivity as the \textit{number of contributions} (i.e., unique revisions or edits made) to article pages (i.e., pages in the article namespace). This means that we exclude considerations like conversation, governance, and interpersonal communication that are also contributions but are made on nonarticle pages.

Finally, we test H4 by looking for changes in \textit{return rate} measured at the user level. First, we follow previous work by \citeauthor{10.1145/2441776.2441873} to break edits into \textit{sessions} defined as ``a sequence of edits made by an editor where the difference between the time at which any two sequential edits are saved is less than one hour''~\cite{10.1145/2441776.2441873}. A user is said to have returned if they made their first edit session to an article page and then made another edit session within 60 days of their first edit session. This is a very low bar for measuring user retention.

Our first three sets of hypotheses are framed in terms of the groups of users who are affected (H\textit{N}a,) and not affected (H\textit{N}b,) and in terms of overall community effects (H\textit{N}c). In practice, this means that we stratify each of the variables described above into outcome measures by groups of users. We operationalize these groups as follows:

\begin{enumerate}
  \item \textbf{IP editors}: Editors who edit content on Wikipedia without an account and whose contribution is credited to their IP address. This group's contributions are subject to FlaggedRevs so this measure is used to test our hypotheses H\textit{N}a.
  \item \textbf{First-time registered editors}: Users who registered an account and made their first edit. This group's contributions are also affected by FlaggedRevs so this measure is used in a second set of tests of hypotheses H\textit{N}a.
  \item \textbf{Returning registered editors}: Users who had contributed at least one previous edit session under a stable identifier. Because each Wikipedia language edition makes choices to determine whether a returning registered editor is ``trusted,'' and because these configurations have been changed over time, it is extremely difficult to determine exactly whether a returning registered editor's work was automatically flagged. That said, it is safe to assume that a large number of contributions made by returning registered editors are not affected by FlaggedRevs because a large majority of contributions to wikis belong to a very small group of veteran contributors \cite{ortega2008inequality}. Because our hypotheses are in support of the null and violation of this assumption will lead to a nonnull effect, this acts as a conservative test of our hypotheses H\textit{N}b.
  \item \textbf{All editors}: All users contributing to a wiki. 
  This group is used to test our hypotheses H\textit{N}c.
\end{enumerate}


\subsection{Analytical Plan}

For our first three hypotheses, we seek to understand the impact of FlaggedRevs at the time it was implemented. To do so, we use Interrupted Time Series (ITS) analysis. ITS is a quasiexperimental research design and is particularly useful when investigators have neither control over the implementation of an intervention nor the power to create treatment and control groups \cite{kontopantelis2015regression}. Our ITS analysis involves constructing a time series regression of our outcome variables (i.e., each of the measures described above in §\ref{sec:outcomes}) and then testing for statistically significant changes in the outcome in the time periods before and after the implementation of the intervention \cite{penfold2013use}. ITS has been used in a number of other social computing studies \cite[e.g.,][]{10.1145/3359265, 10.1145/3134666, 10.1145/3274406}.

ITS relies on a series of assumptions and analytical choices \cite{10.1093/ije/dyw098}. First, it requires a clearly defined intervention with a known implementation date. In our dataset, each wiki has a clear date when FlaggedRevs is enabled. 
The second step is identifying an outcome that is likely to be affected by the intervention. 
For each of our hypotheses, we clearly describe the affected measures above.
Third, ITS requires at least eight observations of an outcome variable at eight different points in time for each period before and after the intervention. Usually, more observations are useful, but having too many data points might have diminishing benefits. We chose to measure activity in each wiki on a monthly basis, over two years, from 12 months before to 12 months after FlaggedRevs was enabled. 

In practice, ITS divides the dataset into two time segments. The first segment comprises rates of the event before the intervention or policy, and the second segment after. ITS applies what is effectively a segmented regression to allow the researcher to test differences in level (i.e., a change in the intercept) as well as change over time (i.e., a change in the slope) associated with the intervention. Segmented regression essentially means that a linear regression model is fit twice within a single model which allows researchers to test for a statistical difference between two time periods.

ITS is typically conducted using a single time series. In our case, we have panel data from 17 different wikis---each with their own trajectory of activity and contributors over time. As a result, we fit wiki-level baseline trends and then seek to estimate the average changes (in both intercept and slope) at the point in time that FlaggedRevs is enabled. The panel data linear regression model that we use for ITS models has the following form:

\begin{equation*}
Y = \beta_0 + {\bm \beta}_{\textbf{\textit{w}}} \textit{\textbf{wiki}} \times \mathit{time} + \beta_1 \mathit{flaggedrev\_on} + \beta_2 \mathit{flaggedrev\_on} \times \mathit{time} + \varepsilon
\end{equation*}

The descriptions of our variables in our model are as follows:

\begin{enumerate}
  \item \textit{Y}: Our dependent variables capturing our outcome measures from some subset of users, as described in the previous section.
  \item \textbf{wiki}: A categorical variable included as a vector of dummy variables indicating the wiki (bold notation indicates that this variable is a vector).
  \item \textit{time}: The month in the study period relative to when FlaggedRevs is enabled. For example, if we measure the outcome of interest one month prior to the day FlaggedRevs is implemented, the time variable would have a value equal to -1.
  \item \textit{flaggedrev\_on}: A dichotomous variable indicating the preintervention period (coded 0) or the postintervention period (coded 1).
\end{enumerate}

We do not report or interpret the coefficients associated with the vector ${\bm \beta}_{\textbf{\textit{w}}}$ which are included only as control variables and which capture different baseline trends for each wiki (e.g., some might be increasing whereas others are decreasing in a given measure). Because we have standardized each measure within each wiki, we do not need to include wiki-level intercept terms because these will all take the same value. Our estimate for $\beta_1$ (associated with \textit{flaggedrev\_on}) estimates the average instantaneous change in intercept immediately following the intervention. Finally, the coefficient $\beta_2$ (associated with $\mathit{flaggedrev\_on} \times \mathit{time}$) indicates the average change in slope following the intervention.

To test H4, we seek to understand how FlaggedRevs was associated with the return rate of new users. Because our return rate measure happens at the level of individual users, we can do better than an overall aggregated rate. Although editors can contribute under different IP addresses, and create different accounts, we treat the first edit made by each unique editor ID as being made by a new user. Our analytical model is a general linear mixed model (GLMM) \cite{mcculloch2014generalized}. 
In particular, we use a logistic binomial regression model with two wiki-level random effects: a random intercept term $u_0$
and a random slope term $u_4$.
These are needed to address issues of repeated measures of users within the same wiki. 
The mixed-effects model that we use is as follows:

\begin{align*}
    \log(\frac{\mathit{returned}}{1-\mathit{returned}}) = &
    (\beta_0 + u_0) +
    \beta_1 \mathit{flaggedrevs\_on} +
    \beta_2 \mathit{first\_edit\_published} +
    \beta_3 \mathit{unregistered}~+ \\
    & (\beta_4 + u_{4})\mathit{time} + 
    \beta_5 \mathit{first\_edit\_published}  \times \mathit{unregistered}~ + \\
    & \beta_6 \mathit{flaggedrevs\_on} \times \mathit{first\_edit\_published}~+  \\
    & \beta_7 \mathit{flaggedrevs\_on} \times \mathit{first\_edit\_published} \times \mathit{unregistered} 
\end{align*}

The descriptions of the variables in our model are as follows:

\begin{enumerate}
  \item \textit{returned}:  A dichotomous variable indicating whether or not an editor made another edit session within 60 days of the first edit session.
  \item \textit{flaggedrevs\_on}: A dichotomous variable indicating whether or not an editor's first edit session was made after the day FlaggedRevs is implemented
  \item \textit{first\_edit\_published}: A dichotomous variable indicating whether or not the final edit in a user's first edit session is published. If the entire session is reverted---as is common---this will revert all of the edits in the session including the final edit. Because revisions can be reverted after they are published, this measure captures whether the editor's first edit session either is never reverted, or it is published (flagged) before being reverted.
  \item \textit{unregistered}: A dichotomous variable indicating whether or not the revision was contributed by someone without an account.
  \item \textit{time}: The month in the study period relative to when FlaggedRevs is turned on. For example, if we measure the outcome of interest one month prior to the day FlaggedRevs is implemented, the time variable would have a value equal to -1.
\end{enumerate}

Our test for H4 is focused on the parameters associated with the interaction terms between our three independent variables (\textit{flaggedrevs\_on}, \textit{first\_edit\_published}, and \textit{unregistered}).
These interactions allow us to understand how the effect of FlaggedRevs on newcomer return rate varies based on whether the user had registered for an account when making their first edit and whether their contribution was visible to the public at some point.

Finally, we inspected the distributions of residuals from each of our regression models to ensure that the key parametric assumptions of our model were not clearly violated. All residuals appeared to be distributed normally and we have placed histograms of these residuals in our online supplementary material.
As we apply multiple statistical tests, we employ a Bonferroni correction to reduce familywise error rate.
In Table \ref{tab:visible-trend-ols}---and all subsequent regression tables---we use an asterisk (*) for p-values that fall under the traditional $\alpha = 0.05$ level and a dagger (\textdagger) next to p-values that fall under the Bonferroni-corrected threshold $(\alpha = 0.05 / 8 = 0.00625)$.
We report results from unadjusted hypothesis tests because many of our results are null and we believe that it is valuable to see that these effects are also null under a lower standard of evidence.

\section{Results}

\begin{figure}
\centering
\includegraphics[width=\columnwidth]{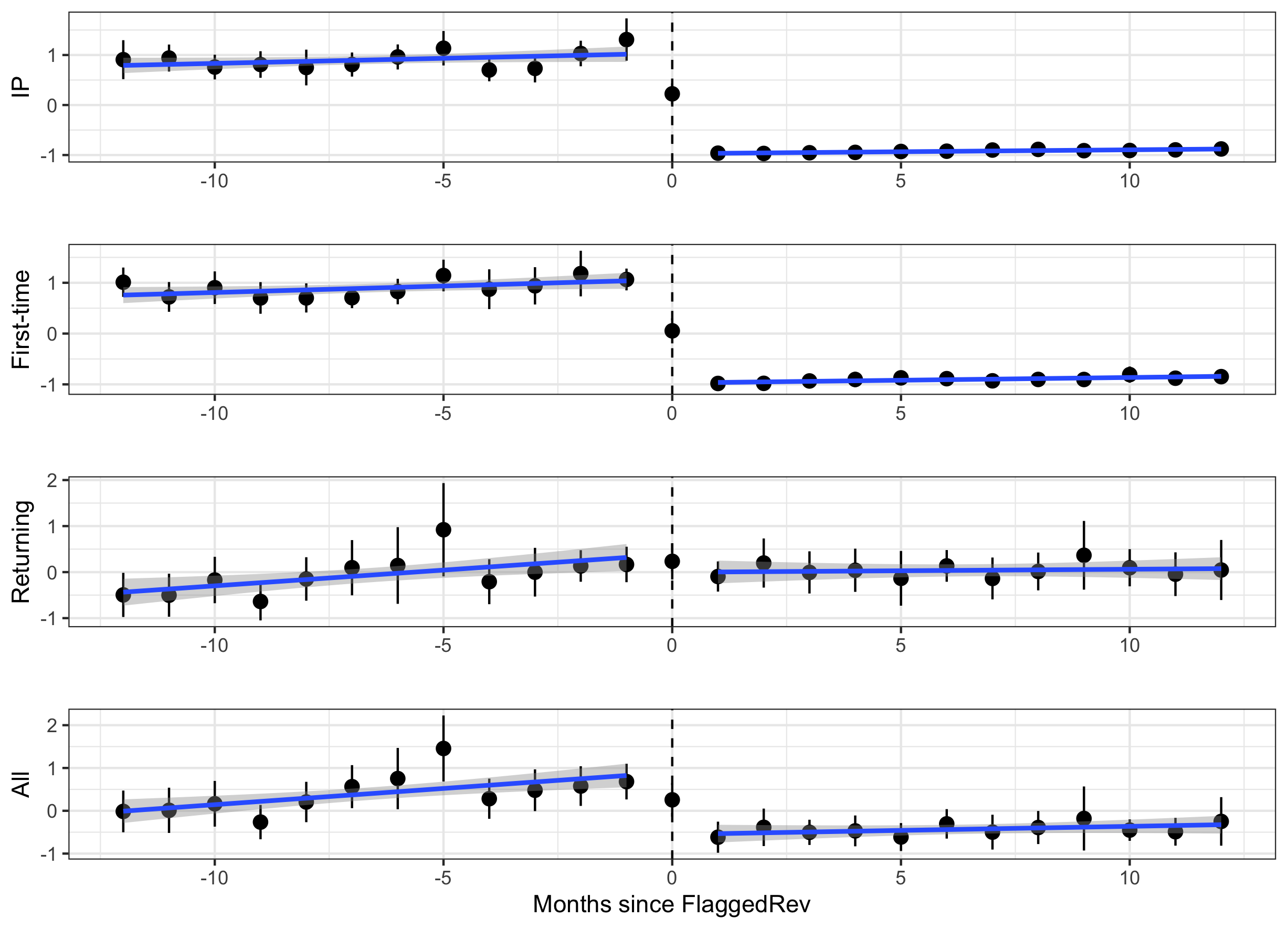}
\caption{Impact of FlaggedRevs on the quantity of visible reverted contributions (in standard deviation units) made by different groups of users. The vertical line indicates the start of the intervention. The regression lines along with confidence bands are fitted using a linear model.}
\label{fig:visible-count}
\end{figure}

To evaluate each of our hypotheses, we first look at a series of visualizations that allow us to visually inspect our data for the hypothesized relationships. Figure \ref{fig:visible-count} shows the data that we use to test the first set of hypotheses (H1) and depicts the trend of each group in the preintervention and postintervention periods. The vertical dashed line indicates the month that FlaggedRevs is deployed. Each data point represents the mean of the number of contributions by an editor group, in standard deviation units. The regression lines and confidence bands are fitted by a simple linear model using \textit{geom\_smooth()} function from ggplot2 in R.  The bands reflect 95\% confidence intervals around the linear model predictions. We choose to clearly differentiate the two periods by excluding the data point at the month of the intervention from the visualized regressions.

\begin{table*}
\caption{Coefficients of the OLS model associated with H1 estimating the number of visible reverted contributions (in standard deviation units) for each group of editors. Models are standardized and include wiki-level fixed effects to control for wiki-level differences in baseline volume and trend.}
\centering
 \begin{tabular}{l l l l} 
\hline
 & Coefficient & Std. Error & p-value \\ [0.5ex] 
 \hline
 \textbf{IP Editors} \\
 \hline
 \textit{flaggedrev\_on} & -1.78 & 0.086 &  <0.001(*)(\textdagger) \\
 $\mathit{flaggedrev\_on}\times\mathit{time}$  &  0.014 & 0.011 &  0.238 \\
 \hline
 \textbf{First-time Editors} \\
 \hline
 \textit{flaggedrev\_on} & -1.759 & 0.095 &  <0.001(*)(\textdagger) \\
 $\mathit{flaggedrev\_on}\times\mathit{time}$  & 0.018 & 0.013 &  0.156 \\
 \hline
 \textbf{Returning Registered Editors} \\
 \hline
 \textit{flaggedrev\_on} & -0.348 & 0.188 & 0.065 \\
 $\mathit{flaggedrev\_on}\times\mathit{time}$  & -0.056 & 0.026 &  0.129 \\
 \hline
 \textbf{All Editors} \\
 \hline
 \textit{flaggedrev\_on} & -1.27 & 0.167 &  <0.001(*)(\textdagger) \\
 $\mathit{flaggedrev\_on}\times\mathit{time}$  & -0.035 & 0.023 &  0.124 \\
 \hline
 \end{tabular}
 \label{tab:visible-trend-ols}
\end{table*}

In Figure \ref{fig:visible-count}, we see evidence of a very clear effect of FlaggedRevs in reducing the number of visible reverted contributions among affected users (H1a) as well as in the overall community (H1c). It is clear that after the intervention, many contributions made by affected users that will eventually be rejected are reverted before they are published. Our statistical tests from our ITS analysis shown in Table \ref{tab:visible-trend-ols} confirm that these effects are statistically significant. As we hypothesized in H1b, the relationship between FlaggedRevs and contribution activities was not statistically significant among returning registered users who appear to be largely unaffected by the moderation system. In general, the results we see are in line with our expectations and provide strong evidence that FlaggedRevs achieves its primary goal of hiding low-quality contributions from the general public.

\begin{figure}
\centering
\includegraphics[width=\columnwidth]{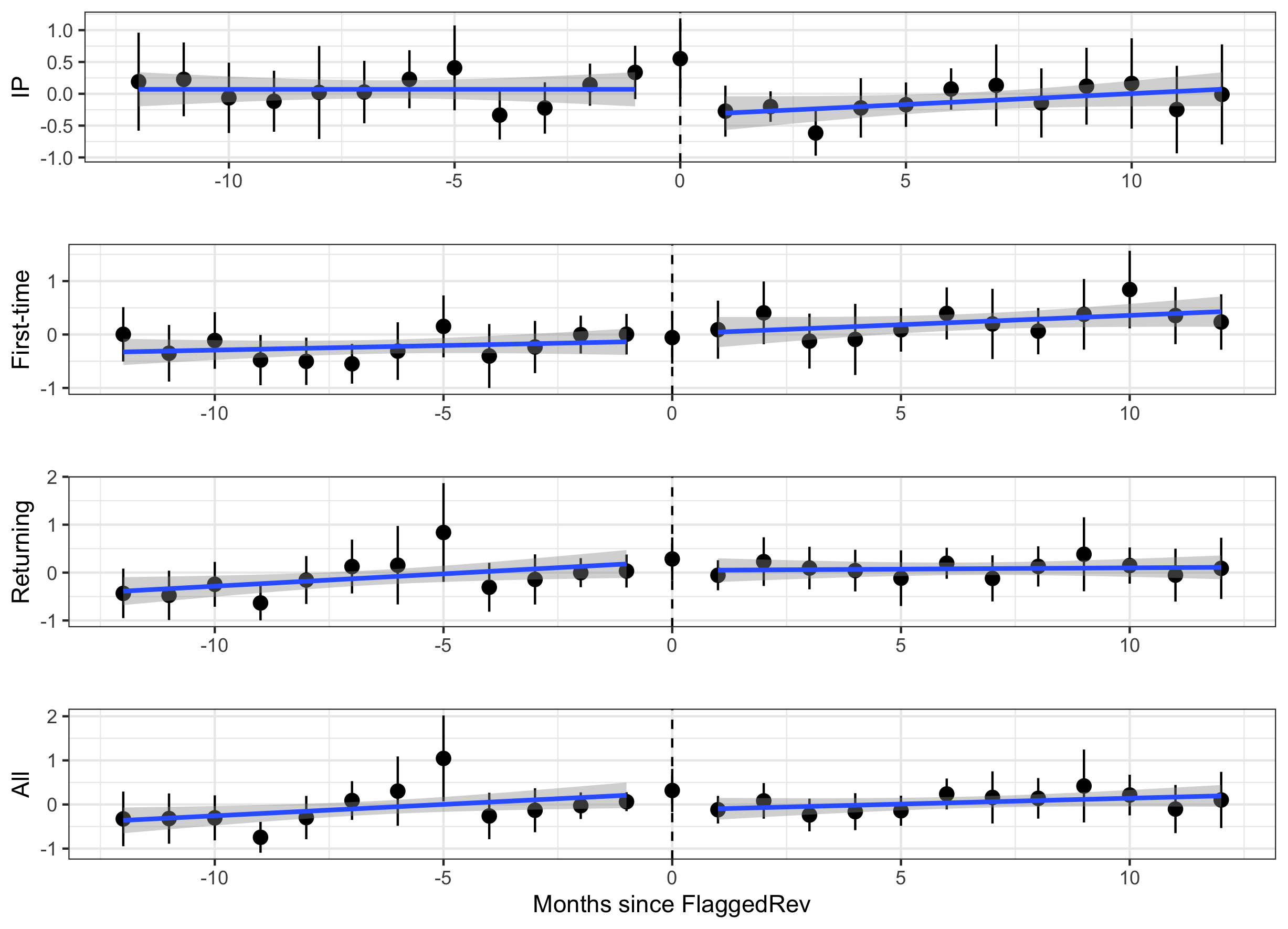}
\caption{Impact of FlaggedRevs on the number of reverted contributions (in standard deviation units) made by different groups of users.}
\label{fig:reverted-count}
\end{figure} 

\begin{figure}
\centering
\includegraphics[width=\columnwidth]{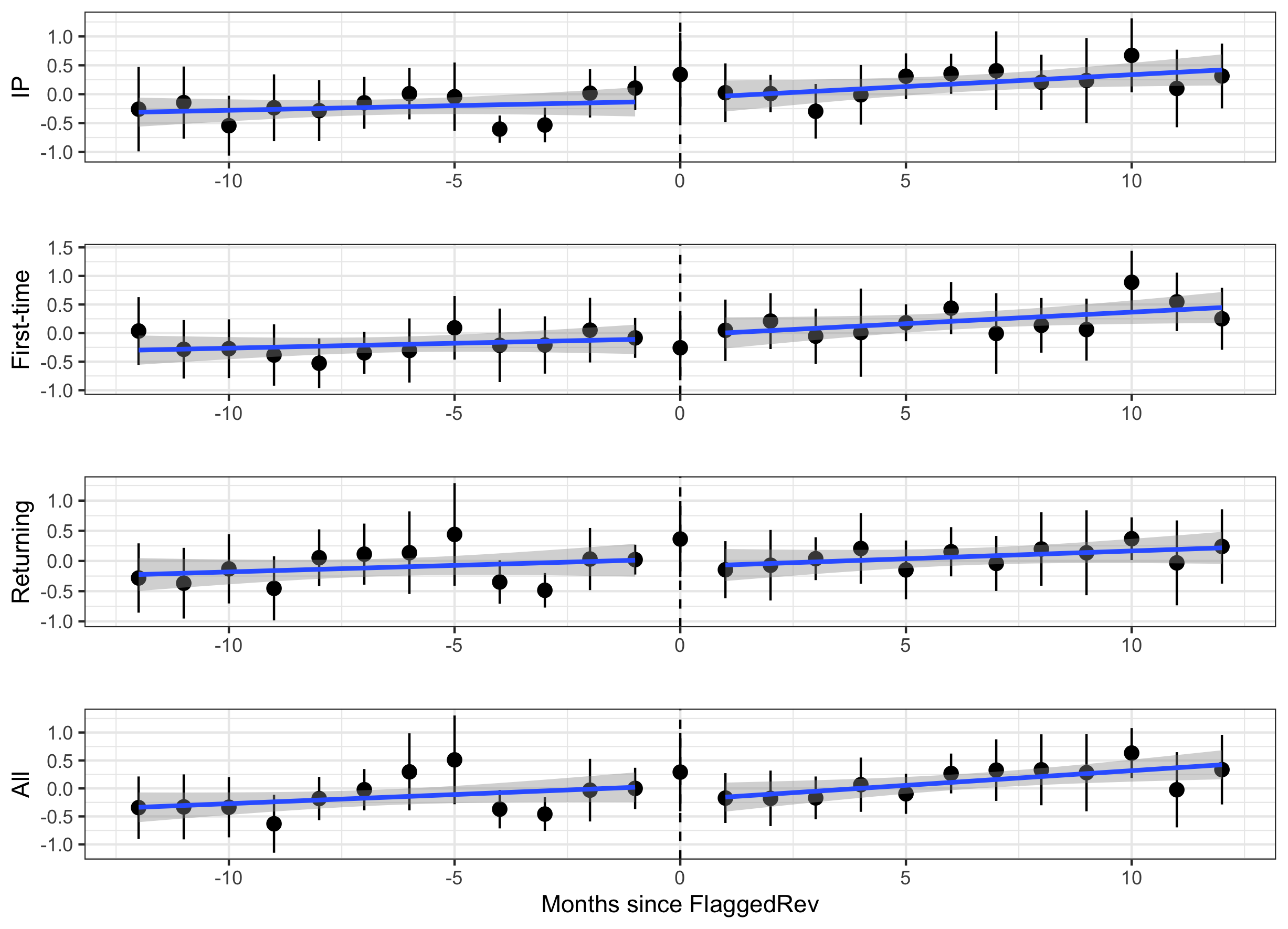}
\caption{Impact of FlaggedRevs on the reverted rate (in standard deviation units) of contributions made by different groups of users.}
\label{fig:reverted-rate}
\end{figure} 

\begin{table*}
\caption{Coefficients of the OLS model associated with H2 estimating the number of reverted contributions for each group of editors. Models are standardized and include wiki-level fixed effects to control for wiki-level differences in baseline volume and trend.}
\centering
 \begin{tabular}{l l l l} 
\hline
 & Coefficient & Std. Error & p-value \\ [0.5ex] 
 \hline
 \textbf{IP Editors} \\
 \hline
 \textit{flaggedrev\_on} & -0.540 & 0.203 &  0.008(*) \\
 $\mathit{flaggedrev\_on}\times\mathit{time}$  &  0.018 & 0.028 &  0.517 \\
 \hline
 \textbf{First-time Editors} \\
 \hline
 \textit{flaggedrev\_on} & 0.019 & 0.213 &  0.640 \\
 $\mathit{flaggedrev\_on}\times\mathit{time}$  & 0.015 & 0.027 &  0.570 \\
 \hline
 \textbf{Returning Registered Editors} \\
 \hline
 \textit{flaggedrev\_on} & -0.200 & 0.202 & 0.322 \\
 $\mathit{flaggedrev\_on}\times\mathit{time}$  & -0.047 & 0.028 &  0.088 \\
 \hline
 \textbf{All Editors} \\
 \hline
 \textit{flaggedrev\_on} & -0.399 & 0.202 &   0.056 \\
 $\mathit{flaggedrev\_on}\times\mathit{time}$  & -0.026 & 0.028 &  0.336 \\
 \hline
 \end{tabular}
 \label{tab:reverted-trend-ols}
\end{table*}

\begin{table*}
\caption{Coefficients of the OLS model associated with H2 estimating the reverted rate for each group of editors. Models are standardized and include wiki-level fixed effects to control for wiki-level differences in baseline volume and trend.}
\centering
 \begin{tabular}{l l l l} 
\hline
 & Coefficient & Std. Error & p-value \\ [0.5ex] 
 \hline
 \textbf{IP Editors} \\
 \hline
 \textit{flaggedrev\_on} & -0.080 & 0.201 &  0.830 \\
 $\mathit{flaggedrev\_on}\times\mathit{time}$  & 0.009 & 0.027 &  0.723 \\
 \hline
 \textbf{First-time Editors} \\
 \hline
 \textit{flaggedrev\_on} & 0.101 & 0.199 &  0.612 \\
 $\mathit{flaggedrev\_on}\times\mathit{time}$  & 0.028 &  0.027 &  0.299 \\
 \hline
 \textbf{Returning Registered Editors} \\
 \hline
 \textit{flaggedrev\_on} & -0.217 & 0.203 &  0.286 \\
 $\mathit{flaggedrev\_on}\times\mathit{time}$  & -0.006 & 0.028 &  0.815 \\
 \hline
  \textbf{All Editors} \\
 \hline
 \textit{flaggedrev\_on} & -0.326 & 0.200 &  0.104 \\
 $\mathit{flaggedrev\_on}\times\mathit{time}$  & 0.012 & 0.027 &  0.664 \\
 \hline
 \end{tabular}
 \label{tab:reverted-rate-gls}
\end{table*}

Our tests for H2 are visualized in Figures \ref{fig:reverted-count} and \ref{fig:reverted-rate} and allow us to visually assess the impact of FlaggedRevs on the volume of low-quality contributions and the average quality among the groups. In the month that FlaggedRevs is installed, we see the number of rejected edits made by IP editors spike significantly. Upon further investigation, we find that when FlaggedRevs was deployed, it retroactively subjected past edits made by the untrusted users to review, regardless of the date that the edit occurred. In this way, the system often required reviewers to assess a large volume of past contributions and resulted in a large number of reverts. After the deployment, the number of reverted contributions among untrusted IP editors quickly drops, before gradually rising. The results from statistical tests in Table \ref{tab:reverted-trend-ols} only confirms a significant impact of FlaggedRevs on the number of reverted contributions made by IP editors before applying the Bonferroni correction, but not after.
Surprisingly, we do not find the same effect on our second group of affected editors (i.e., first-time registered editors). Although the number of rejected contributions shown in Figure \ref{fig:reverted-count} appears to increase gradually, this trend is consistent with the preintervention trend, suggesting that FlaggedRevs may not play a role in the trend. We also do not see strong evidence of a major effect of FlaggedRevs on either unaffected editors or the community overall.

The impact of FlaggedRevs on the revert rate of contributions made by different groups of users shown in Figure \ref{fig:reverted-rate} and Table \ref{tab:reverted-rate-gls} tells a similar story. We see neither a significant immediate change nor a change in trajectory associated with the intervention. Although we do see a slight uptick in the revert rate during the month that FlaggedRevs is deployed, the rate quickly goes back down to the mean level before gradually increasing over time. Our overall results for H2 reflect a consistent null result. We find little evidence of the prepublication moderation system having a major impact on the quality of contributions. Thus, we cannot conclude that FlaggedRevs alters the quantity or quality of newcomers’ contributions.

\begin{figure}
\centering
\includegraphics[width=\columnwidth]{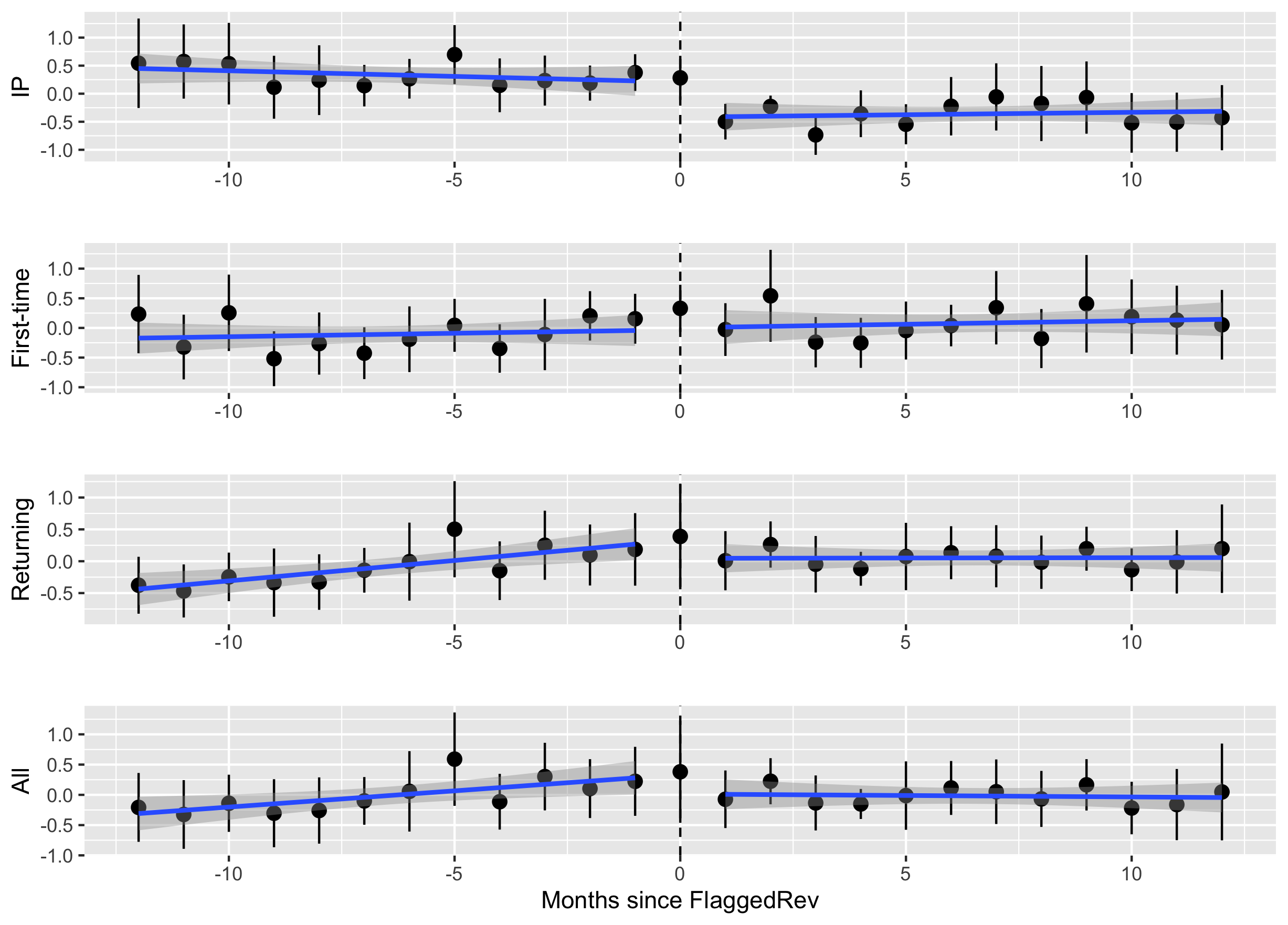}
\caption{Impact of FlaggedRevs on the number of contributions (in standard deviation units) among different groups of users.}
\label{fig:edit-trend}
\end{figure} 

\begin{table*}[h]
\caption{Coefficients of the OLS model associated with H3 estimating the number of edits (in standard deviation units) made by each group of editors. Models are standardized and include wiki-level fixed effects to control for wiki-level differences in baseline volume and trend.}
\centering
 \begin{tabular}{l l l l} 
\hline
 & Coefficient & Std. Error & p-value \\ [0.5ex] 
 \hline
 \textbf{IP Editors} \\
 \hline
 \textit{flaggedrev\_on}  & -0.626 & 0.209 & 0.003(*) \\
 $\mathit{flaggedrev\_on}\times\mathit{time}$ & 0.0287 & 0.028 &  0.3120 \\
 \hline
 \textbf{First-time Editors} \\
 \hline
 \textit{flaggedrev\_on} & <0.001 & 0.030 &  0.994 \\
 $\mathit{flaggedrev\_on}\times\mathit{time}$  & 0.0348 & 0.222 &  0.876 \\
 \hline
 \textbf{Returning Registered Editors} \\
 \hline
  \textit{flaggedrev\_on}  & -0.285 & 0.192 & 0.139 \\
  $\mathit{flaggedrev\_on}\times\mathit{time}$ & -0.063 & 0.026 & 0.067 \\
 \hline
 \textbf{All Editors} \\
 \hline
  \textit{flaggedrev\_on}  & -0.320 & 0.209 & 0.200 \\
  $\mathit{flaggedrev\_on}\times\mathit{time}$ & -0.064 & 0.028 & 0.062 \\
 \hline
 \end{tabular}
 \label{tab:edit-trend-gls}
\end{table*}

Regarding our H3 hypotheses, Figure \ref{fig:edit-trend} once again shows two different stories for the two editor groups that are affected by FlaggedRevs. The deployment of the system appears to be associated with an immediate decline in the number of contributions made by IP editors, but not enough to reject the null hypothesis. This result holds \textit{before} the Bonferroni correction is applied, but not after. The ITS regression results shown in Table \ref{tab:edit-trend-gls} confirm that the visually apparent effect described above is significant.
Once again, we do not see a similar effect within the first-time registered editor group also affected by FlaggedRevs. Although the system did not appear to cause an immediate change in the number of contributions among the returning registered editor group and the all-editor group, it does appear associated with a change in trajectory compared to the preintervention period. That said, it does not correspond to negative growth (the postintervention edit trend remains fairly flat and hovers near the mean level). Overall, we see the deployment of the prepublication discouraged the participation of the group of editors with the lowest commitment and most targeted by the additional safeguard, but not the other groups. 

\begin{figure}
\centering
\includegraphics[width=0.7\columnwidth]{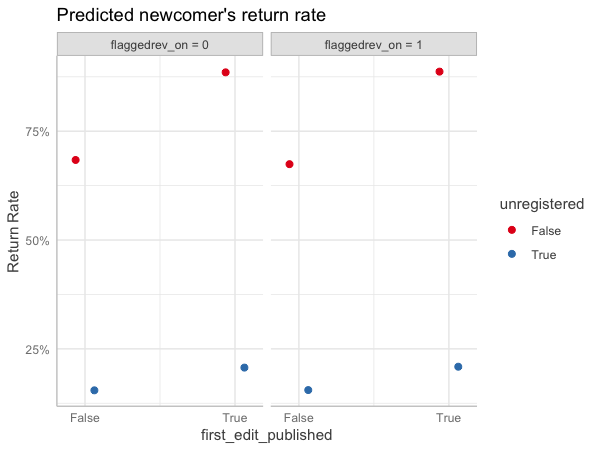}
\caption{Predicted return rate from our model four prototypical newcomers from our model testing H4. These predicted values help interpret the three-way interaction between different independent variables on newcomer return rate on wikis that implemented FlaggedRevs.}
\label{fig:glmer-interaction}
\end{figure}

Finally, our test of H4 is reported in Table \ref{tab:glmer-model} which reports the estimates from our GLMM estimating newcomer return rate. We find that although FlaggedRevs did negatively affect the return rate of newcomers in a way that was statistically significant, the size of this effect is extremely small. 
Because three-way interactions can be difficult to interpret, we report a range of predicted values from our models and visualize these results in Figure \ref{fig:glmer-interaction}.
Our models suggest that before FlaggedRevs was enabled,  67\% of newcomers return to make a second edit session within 620 days of their first session when that first edit session was not published and when they registered for an account. Our model suggests that the return rate is reduced by 2\% after FlaggedRevs is enabled.


\begin{table*}
\caption{Estimated values of the multilevel logistic model estimating the likelihood of newcomer return rate. In addition to the reported parameters, the model includes a wiki-level random intercept term.}
\centering
 \begin{tabular}{l l l l} 
\hline
 & Estimated Value & Std. Error & p-value \\ [0.5ex] 
 \hline
 (Intercept) & 0.771 & 0.159 & <0.001 (*)(\textdagger) \\
 \textit{flaggedrevs\_on} & -0.044 & 0.020 & 0.031 (*) \\
  \textit{first\_edit\_published} & 1.273 & 0.014 & <0.001 (*)(\textdagger) \\
  \textit{unregistered} & -2.470 & 0.016 & <0.001 (*)(\textdagger) \\
  \textit{time} & -0.015 &  0.006 & 0.024 (*) \\
 \hline
  \textit{first\_edit\_published} $\times$  \textit{unregistered} & -0.915 & 0.017 & <0.001 (*)(\textdagger) \\
 \textit{flaggedrevs\_on} $\times$  \textit{first\_edit\_published} & 0.043 & 0.022 & 0.053 \\
 \textit{flaggedrevs\_on} $\times$  \textit{first\_edit\_published} $\times$  \textit{unregistered} & -0.056 &  0.024 & 0.020 (*) \\
 \hline
 Number of observations: 1,972,861 \\
 \hline
 \end{tabular}
 \label{tab:glmer-model}
\end{table*}

Many of our findings are null effects. In frequentist statistical analyses, null effects are not evidence of noneffects in that they could be caused by large variation and/or measurement error, small sample sizes, small real effect sizes, or some combination of these factors.  Although many of our estimates are not statistically significant, the large majority of these nonsignificant parameter estimates are near zero. In other words, we would fail to reject the null hypothesis even if our standard errors were much smaller---in some cases several orders of magnitude smaller. On the other hand, some of our standard errors are 0.1-0.5 standard deviations in magnitude, suggesting that there is meaningful uncertainty around our estimates.

To provide a reasonable bound for the size of likely effects, we conducted a \textit{post hoc} power analysis using bootstrapping and simulation.
For each null result, we first subtracted out any estimated effect from our dataset so that the parameter estimate in question was precisely 0. Next, we generated sample datasets in which an effect was known to be present by adding back effects of a predetermined size.  Finally, we conducted bootstrap-based simulations by sampling with replacement from our dataset 2,000 times to create equally sized new datasets. Using each of these 2,000 bootstrapped datasets, we fit a new model and determined whether our effect was, or was not, statistically significant. Finally, we counted the proportion of statistically significant results to estimate our statistical power. 
Our wiki-level and user-level analyses were bootstrapped at the wiki- and user-level, respectively.
A table of full results from these analyses is reported in our online supplement.

Regarding H1, H2, and H3, our simulations suggest that our model would be able to observe a true effect of 0.4 standard deviations associated with our independent variable \textit{flaggedrevs\_on} at least 80\% of the time. For our parameter associated with $\mathit{flaggedrev\_on}\times\mathit{time}$, our model would be able to observe a true effect of 0.5 standard deviations at least 80\% of the time. Regarding H4, our model would be able to detect an effect size of 0.2 standard deviation in any parameter 95\% of the time.
The body of evidence suggests that if any effects of FlaggedRevs on our outcomes exist, either as a difference in the instantaneous effect or a change in slope over time, it is likely less than half a standard deviation difference. These results reflect the fact that there is substantial variation in effects across wikis but that our best estimates, especially for user-level effects, suggest that any real effects of the system on any of our outcomes are likely moderate in size at most.

\section{Discussion}

Our results suggest that prepublication moderation systems like FlaggedRevs may have a substantial upside with relatively little downside. If this is true, why are a tiny proportion of Wikipedia language editions using it? Were they just waiting for an analysis like ours? In designing this study, we carefully read the Talk page of FlaggedRevs.\footnote{\url{https://meta.wikimedia.org/wiki/Talk:Flagged\_Revisions}} Community members commenting in the discussion agreed that prepublication review significantly reduces the chance of letting harmful content slip through and being displayed to the public. Certainly, many agreed that the implementation of prepublication review was a success story in general---especially on German Wikipedia. Others pointed out that German Wikipedia has tens of thousands of volunteers who actively monitor and review contributions in a timely manner. As a result, nearly 100\% of contributions under review are approved or rejected by moderators in less than 24 hours.\footnote{\url{https://de.wikipedia.org/wiki/Spezial:Sichtungsstatistik}} 

However, the same discussion also reveals that the success of German Wikipedia is not enough to convince more wikis to follow in their footsteps. From a technical perspective, FlaggedRevs' source code appears poorly maintained.\footnote{\url{https://phabricator.wikimedia.org/T66726\#3189794}} Additionally, FlaggedRevs is designed to be highly customizable with many options. This, along with the time commitment required to review changes, makes the system---and likely many other prepublication review systems that must be configured with a similar degree of granularity---difficult to adopt. 

FlaggedRevs itself suffers from a range of specific limitations. For example, the FlaggedRevs system does not notify editors that their contribution has been rejected or approved. Some users with review rights complained that ``[u]ser interface is not made for checking all edits in real time.''\footnote{\url{https://meta.wikimedia.org/wiki/Requests\_for\_comment/Flagged\_revisions\_deployment}}
Indeed, users often pointed to the unintuitive user interface as one of the reasons why wikis using FlaggedRevs have struggled to get enough members to review unflagged revisions.
Moreover, the absence of detailed exchange among language editions on the effectiveness of FlaggedRevs---like our study---meant that Wikipedia administrators had little incentive to review the FlaggedRevs code base to maintain and improve it.
Since April 2017, requests for deployment of the system by other wikis have been paused by the Wikimedia Foundation indefinitely.\footnote{\url{https://phabricator.wikimedia.org/T66726\#3189794}} 
Despite these problems, our findings suggest that the system kept low-quality contributions out of the public eye and did not deter contributions from the majority of new and existing users. Our work suggests that systems like FlaggedRevs deserve more attention.

How should platforms decide whether and how to deploy a prepublication moderation system like FlaggedRevs? For platforms such as Wikipedia, Reddit, and Fandom, there exist many subcommunities within the platform with a high degree of autonomy. Due to the vastly different sizes, norms, and languages spoken across subcommunities \cite{10.1145/3274301}, content moderation is mostly conducted by volunteers. Many platforms use a ``community-reliant approach'' to content moderation, which means that the parent organization only sets up overarching norms and standards that are then enforced by subcommunity administrators \cite{caplan_2018, mcgillicuddy2016controlling}. We believe that prepublication moderation should be deployed in this way as well and that moderators who know their communities best should make the important decisions regarding who and what should be held for vetting and who should be allowed to vet.

Prepublication moderation may act as both an alternative and/or a complement to other types of moderation systems.
Any decision to deploy a system like FlaggedRevs should compare the system to realistic alternative approaches. These approaches often include wholesale blocking of groups of users and automated forms of postpublication review, either of which may lead to negative consequences.
For example, in situations where automated moderation carries the risk of diminishing human connections between community members, it has often been found to be inaccurate, biased, and lacking in transparency and accountability \cite{10.1007/978-3-319-67256-4_32, gorwa2020algorithmic, gillespie2020content, 10.1145/3338243, 10.1145/3415238}. Mandatory prepublication review of content might be an improvement over automatic deletion of potentially problematic content in some cases.

Similarly, we believe that our work suggests that---at a minimum---a streamlined version of systems like FlaggedRevs could serve as a path for reviewing contributions from populations of contributors that are currently deemed too ``high risk'' to contribute to peer production systems at all. For example, previous research by our team has shown that contributions from anonymity-seeking Tor users (who are currently blocked from contributing to Wikipedia altogether) have been a source of substantial value in the past \cite{chau_tor_ieee}. Although anonymous activities can be hard to govern, they can ``encourage expressiveness and interaction among users'' \cite{10.1145/2818048.2820081}. The implementation of a prepublication moderation system could prove effective in governing the content made by these less privileged users, fostering a culture that allows diverse voices and open discussion, instead of outright excluding them due to perceived risk.

\section{Limitations}

One limitation with our analysis stems from the heterogeneity in the communities deploying FlaggedRevs. 
For example, wikis of different sizes also have different numbers of editors with review rights leading to vastly different average review times. For example, German Wikipedia currently has 19,994 users with review rights, and the median review delay for edits by users without accounts is two hours.\footnote{\href{https://de.wikipedia.org/wiki/Spezial:Sichtungsstatistik}{German Wikipedia Patrol Statistics} (Archived at \url{https://perma.cc/PL7G-M2ZE})} Meanwhile, Russian Wikipedia has 2,422 users with review rights, and the median review delay for similar edits is more than 13 days.\footnote{\href{https://ru.wikipedia.org/wiki/\%D0\%A1\%D0\%BB\%D1\%83\%D0\%B6\%D0\%B5\%D0\%B1\%D0\%BD\%D0\%B0\%D1\%8F:\%D0\%A1\%D1\%82\%D0\%B0\%D1\%82\%D0\%B8\%D1\%81\%D1\%82\%D0\%B8\%D0\%BA\%D0\%B0_\%D0\%BF\%D1\%80\%D0\%BE\%D0\%B2\%D0\%B5\%D1\%80\%D0\%BE\%D0\%BA}{Russian Wikipedia Patrol Statistics} (Archived at
\url{https://perma.cc/B3EV-82BC})}
Our null result may be attributable, at least in large part, to high variation in the size of the estimated relationships. Simply put, the experience of prepublication review may vary enormously. 

A second limitation stems from our measure construction for testing H4 and H1-3b. A threat to validity stems from the fact that our measure includes both affected and nonaffected users. Unfortunately, because we do not have access to wiki-level configuration data on FlaggedRevs or the ability to match IP addresses to users, we cannot easily identify which particular users were affected. Because the large majority of contributions to Wikipedia are from a very small group of veteran editors, we believe that this is unlikely to be driving our results in our user-level analysis. Although this means that hypothesis tests in H4 are likely conservative estimates, our estimates for H4 should be interpreted with some caution.

A final limitation of our study concerns its generalizability. Ultimately, like all studies of a single platform or website and the large majority of empirical studies in social computing, we cannot know whether or how our results will generalize to other settings. We have three reasons to believe that our work is a valuable contribution to social computing. First, we gain some confidence from the fact that we are able to report the effect of the system across 17 communities, not just one. In the ways discussed above, these communities vary along a range of dimensions including size, language, and culture. 
Second, there are many other communities in different platforms that could deploy similar prepublication moderation systems. For example, the Fandom platform hosts hundreds of thousands of wikis running MediaWiki and could deploy FlaggedRevs directly. Certain subcommunities on Reddit also use the \textit{user flair} tagging system to identify new users and may require their contributions to be reviewed before being publicly displayed. Since 2017, Quora has subjected contributions by unregistered users to prepublication moderation.\footnote{\url{https://techcrunch.com/2017/02/10/qa-site-quora-clamps-down-on-anonymity-will-review-content-before-publishing-restricts-actions/}} Finally, we believe that Wikipedia's size and importance make our results valuable even if our work is limited in its generalizability. Wikipedia is a top-10 website that exists in more than 300 languages and more than 90\% of Wikipedia editions---including English Wikipedia---have not yet deployed FlaggedRevs. Our results provide evidence that can inform ongoing conversations regarding decisions to employ prepublication moderation in these settings and beyond.

\section{Conclusion}

This study aimed to measure the risks, benefits, and unintended effects associated with a substantial change to the moderation systems within a set of 17 communities. We did so by presenting a case study of FlaggedRevs, a system designed by German Wikipedia and deployed by other Wikipedia language editions.
First, we sought to understand if the deployment of FlaggedRevs did what it was designed to do. We found that in this regard, it was an unambiguous and unmitigated success. By adding prepublication moderation, the Wikipedia language editions in our sample kept a large portion of vandalism and other low-quality contributions by untrusted users from ever being seen by the public.

Contrary to our hypothesis, we did not find strong evidence of any meaningful long-term change in contribution quality. This suggests that communities that change their content moderation from postpublication to prepublication to discourage poor-quality contributions from ever occurring may not see the relief they seek.
Furthermore, we found that FlaggedRevs does not appear to have had a major impact on newcomer retention; although the effect was negative and statistically significant, it was small in size. 
This suggests that communities concerned about driving away newcomers if they shift from postpublication to prepublication may find the impact is not so dramatic.

Although the null results in our analysis can be difficult to convincingly interpret, we believe that findings like ours can play an important role in contributing to social computing theory and systems. These communities that we report on were able to successfully deploy prepublication moderation system as an additional quality control measure to eliminate inappropriate content without substantial decreases in newcomer retention. FlaggedRevs could provide valuable insights and lessons for online platforms that wish to combat vandalism without sacrificing openness to outsiders, newcomers, or anonymity seekers.


\begin{acks}
We owe a particular debt of gratitude to Tilman Bayer, Andrea Forte, Charles Kiene, and Shruti Sannon who all contributed enormously valuable feedback. We also thank the anonymous reviewers and program committee members at CSCW for their helpful guidance. The  creation  of  our dataset  was  aided  by the  use  of  advanced  computational,  storage,  and  networking infrastructure provided by the Hyak supercomputer system at the  University  of  Washington.  This  work  was  supported  by the  National  Science  Foundation  (awards CNS-1703736  and CNS-1703049).
\end{acks}

\section*{Data and Code}

A replication dataset including data, code, and other supplementary material has been placed in the Harvard Dataverse archive and is available at: \url{https://doi.org/10.7910/DVN/G1YFLE}










\bibliographystyle{ACM-Reference-Format}
\bibliography{main}

\end{document}